\documentclass[a4paper,onecolumn,12pt,assc]{IEEEtranASSC}
\ifCLASSINFOpdf
   \usepackage[pdftex]{graphicx}
\else
\fi
\begin{document}
%
\title{An Impact Crater in Palm Valley, Central Australia?}


\author{\IEEEauthorblockN{Duane W. Hamacher\IEEEauthorrefmark{1},
Andrew Buchel\IEEEauthorrefmark{2}, Craig O'Neill\IEEEauthorrefmark{2},
Tui R. Britton\IEEEauthorrefmark{3}}

\IEEEauthorblockA{\IEEEauthorrefmark{1}
Department of Indigenous Studies and the Research Centre for Astronomy, Astrophysics \& Astrophotonics, Macquarie University, NSW, 2109, Australia}

\IEEEauthorblockA{\IEEEauthorrefmark{2}
Department of Earth \& Planetary Science and the Centre for Geochemical Evolution \& Metallogeny of Continents, Macquarie University, NSW, 2109, Australia}

\IEEEauthorblockA{\IEEEauthorrefmark{3}
Department of Physics \& Astronomy and the Research Centre for Astronomy, Astrophysics \& Astrophotonics, Macquarie University, NSW, 2109, Australia}}

\maketitle

\begin{abstract}
We explore the origin of a $\sim$280~m wide, heavily eroded circular depression in Palm Valley, Northern Territory, Australia using gravity, morphological, and mineralogical data collected from a field survey in September 2009.  From the analysis of the survey, we debate probable formation processes, namely erosion and impact, as no evidence of volcanism is found in the region or reported in the literature.  We argue that the depression was not formed by erosion and consider an impact origin, although we acknowledge that diagnostics required to identify it as such (e.g. meteorite fragments, shatter cones, shocked quartz) are lacking, leaving the formation process uncertain.  We encourage further discussion of the depression's origin and stress a need to develop recognition criteria that can help identify small, ancient impact craters.  We also encourage systematic searches for impact craters in Central Australia as it is probable that many more remain to be discovered.
\end{abstract}


\section*{Introduction}

Unusual circular depressions can result from a number of geologic processes, including volcanism, erosion/collapse, and impact cratering.  The identification of those formed by impacts is useful in providing data necessary to constrain models of impact physics, rock and debris mechanics, crater population distributions, and estimates of the meteoroid influx \cite{IEEEhowto:grieve}\cite{IEEEhowto:Stothers}\cite{IEEEhowto:Nurmi}, especially regarding smaller meteoroids ($<$~20~m) \cite{IEEEhowto:Silber}.  The ancient, arid deserts of Central Australia, being Mesozoic or greater in age \cite{IEEEhowto:Twidale}\cite{IEEEhowto:Ollier}, provide good conditions for the preservation of impact structures. Of the 28 confirmed craters in Australia to date, with an additional 14 probable or proposed structures awaiting confirmation \cite{IEEEhowto:Haines}\cite{IEEEhowto:Iasky}\cite{IEEEhowto:Glikson}, only five are located in the Central Australian region (Gosse's Bluff, Henbury, Boxhole, Kelly West, and Amelia Creek).  Of these five, only two have ages $<$~140~Ma (Boxhole and Henbury), both of which are only a few thousand years old \cite{IEEEhowto:Haines}.  Current estimates of the meteoroid influx rate \cite{IEEEhowto:Collins} suggest more craters remain to be discovered in Central Australia.  

Publicly accessible high-resolution mapping technology, such as Google Earth, is making the search for these craters substantially easier.  But it is difficult to discriminate between circular features that are formed by terrestrial processes and those formed by impacts using satellite imagery alone.  To identify an impact origin, French and Koeberl \cite{IEEEhowto:French} provide a useful treatise on the convincing and unconvincing evidence for impact structures.  They list eight Class~A diagnostic indicators that are formed only from impact events (such as meteorites, shatter cones, and shocked quartz) and ten Class~B non-diagnostic indicators that are caused by impact events as well as other geologic processes (Table~1).

\begin{table}
 \centering
 \begin{minipage}{15cm}
 \label{tab:diagnostics} 
\caption{Shock-produced deformation effects: diagnostic (Class~A) and non-diagnostic (Class~B) indicators, taken from French \& Koeberl \cite{IEEEhowto:French}.}
\begin{tabular}{ll}
\hline
 Class A (Diagnostic) Indicators				&	Class B (non-Diagnostic) Indicators	\\
\hline	
	Preserved meteorite fragments				&	Circular morphology\\
	Chemical and isotopic projectile signatures	&	Circular structural deformation\\
	Shatter cones							&	Circular geophysical anomalies\\
	High-pressure (diaplectic) mineral glasses	&	Fracturing and brecciation\\
	High-pressure mineral phases				&	Kink banding in micas\\
	High-temperature glasses and melts		&	Mosaicism in crystals\\
	Planar fractures (PFs) in quartz 			&	Pseudotachylite/breccias\\
	Planar deformation features (PDFs) in quartz 	&	Igneous rocks and glasses\\
										&	Spherules and microspherules\\
										&	Other problematic criteria		\\		
\hline
\end{tabular}
\end{minipage}
\end{table}

It is difficult to identify diagnostic criteria for small (D~$<$~500~m) impact structures, as diagnostics such as shocked quartz, high pressure melt, and shatter cones require a minimum pressure that is not generally achieved, whilst other diagnostics, such as meteorite fragments and ejecta, may be destroyed, degraded, or obscured by erosion, particularly for older impacts.

\begin{figure}
\centering
\includegraphics[width=16cm]{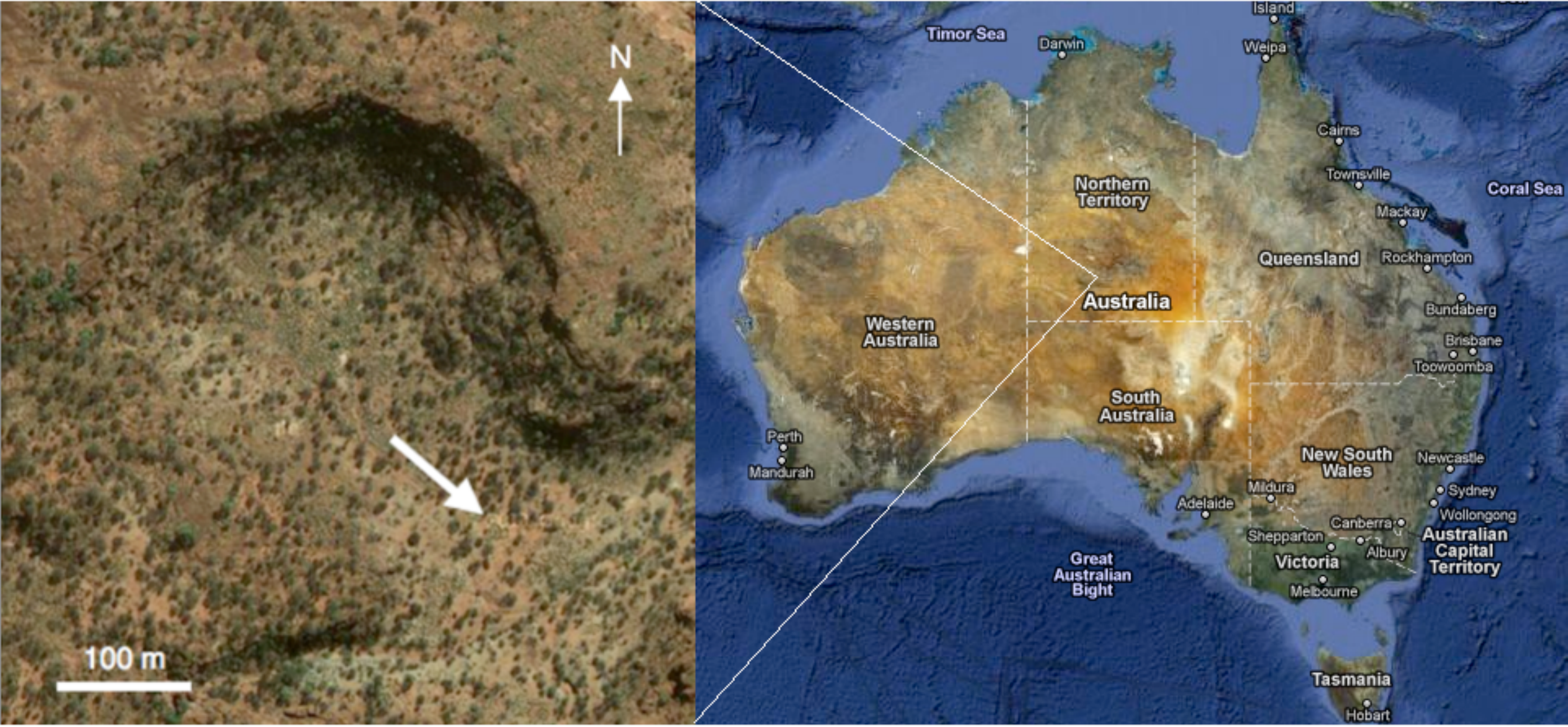}
\caption{A Google Maps image (2009) of the Palm Valley depression, showing its location in the Northern Territory.  The white arrow indicates the outflow channel to the southeast.}
\label{fig:1}
\end{figure}

This paper presents results of a preliminary field survey that explores the origins of a small $\sim$280~m wide depression (24$^{\circ}$ 03$^{\prime}$ 05$^{\prime \prime}$ S, 132$^{\circ}$ 42$^{\prime}$ 33$^{\prime \prime}$ E, Fig.~\ref{fig:1}) located in Palm Valley, Northern Territory, Australia identified using Google Earth.  This study utilises the French and Koeberl criteria for diagnosing impact craters but emphasises the need to identify convincing criteria for small, eroded craters.

\section{Geologic History of Palm Valley}

Palm Valley lies $\sim$125~km southwest of Alice Springs in the Krichauff Ranges within the Finke Gorge National Park, 13~km southwest of Hermannsburg.  The region is dominated by an extensive outcropping of Devonian Hermannsburg Sandstone, a major unit of the Pertnjara Group.  The Pertnjara Group is the youngest depositional section of the Amadeus Basin, an extensive intracratonic sedimentary basin of Neoproterozoic to Palaeozoic age extending over Central Australia, covering an area of $\sim$170,000~km$^2$ \cite{IEEEhowto:Wischusen}.

The Hermannsburg Sandstone is indurated, poorly sorted lithic-felspathic quartz sandstone with Wentworth grain sizes ranging from fine (125-250~$\mu$m) to very course (1-2~mm) that is deposited in fining upward cycles to a thickness of 1.3~km.  At outcroppings, cream calcrete capping can be commonly observed, as well as joint/fracture lines with calcrete joint infilling.  Iron staining is also a predominant feature of this unit, giving it a characteristic red to brown colour \cite{IEEEhowto:Jones72}\cite{IEEEhowto:Warren}.  Deposition of Hermannsburg Sandstone began in the Middle to Upper Devonian during the Pertnjara Movement event and continued until the Late Devonian Alice Springs Orogeny.  This coincides with broad folding of the entire sequence and the start of deposition of the overlying Brewer Conglomerate.  The depositional environment is thought to be that of an extensive braid-plain river system, inferred using composition, grain size, and lack of siltstone lens as evidence \cite{IEEEhowto:Jones91}.  Furthermore, the lack of coarse scale detritus indicates low relief and/or distant source rock.

Palm Valley lies in the Finke Gorge system, part of the ephemeral Finke River.  The antiquity of the river is argued to range from 300-400~Ma on the basis of deeply incised meanders \cite{IEEEhowto:Pickup}. These meanders form only on moderately level plains, suggesting they predate the uplift on the James Ranges, which are associated with the Alice Springs orogeny.  However, large portions of the southern section of the river were inundated by an inland sea during the Mesozoic and so are certainly younger \cite{IEEEhowto:Wells}. In either case projected erosion rates for the river system are extraordinarily low.

\section{Site Survey}

\begin{figure}[h]
\centering
\includegraphics[width=13cm]{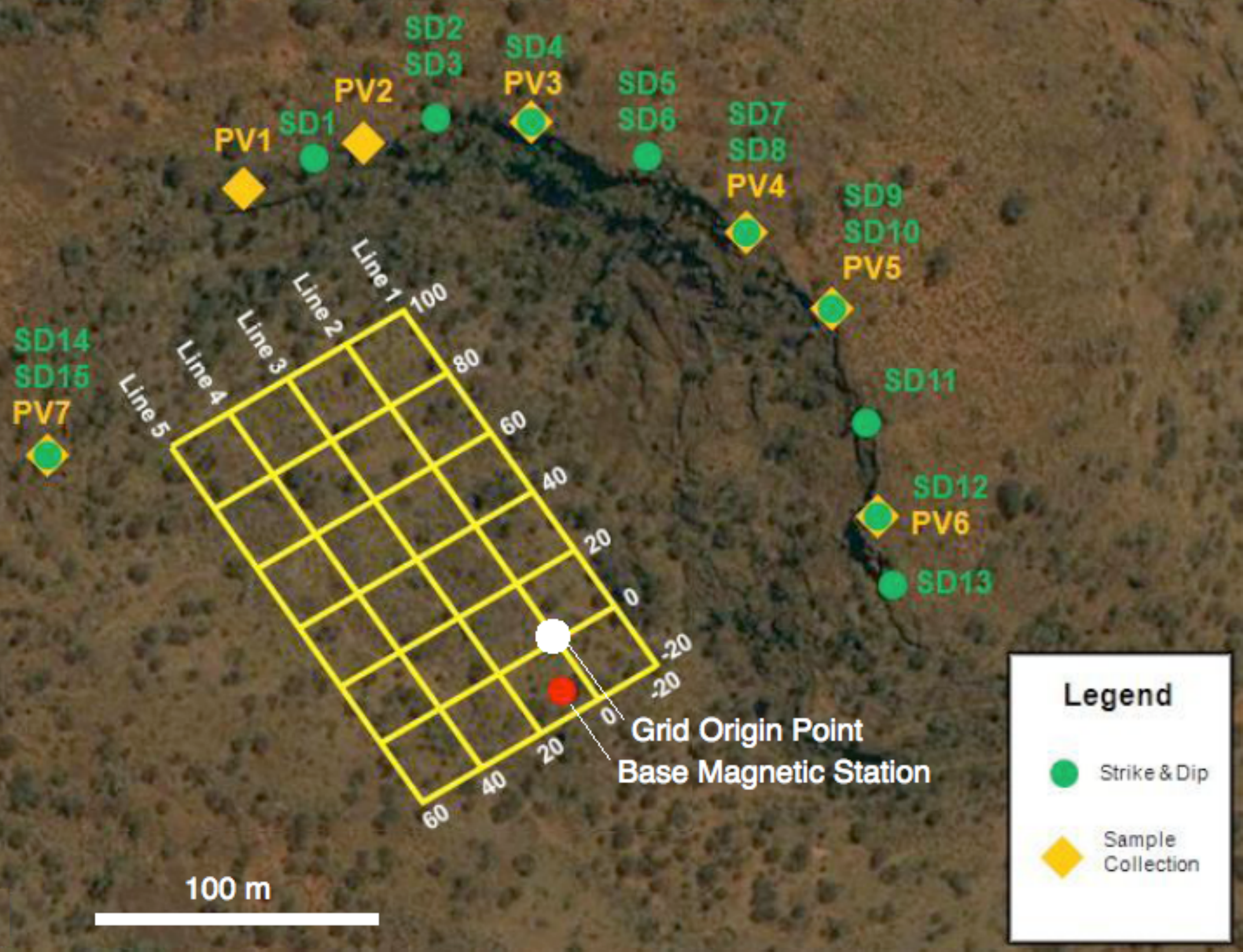}
\caption{Map of the depression's floor showing grid points from which gravity data were taken and points on the depression's edge from which strike and dip measurements and rock samples were taken.  Image taken from Google Earth.  North is to the top.}
\label{fig:2}
\end{figure}

A geophysical and topographical survey of the depression was conducted on 8-9 September 2009.  The depression's floor was topographically mapped and gravity and magnetic data were collected.  Unfortunately, the contrast in magnetic susceptibility between the sedimentary infill and the sandstone was insufficient to provide a significant response, so the results of the magnetics will not be discussed further.  The gravity and topography surveys were conducted on a grid with an origin point peg (0, 0) at 24$^{\circ}$ 03$^{\prime}$ 5.0$^{\prime \prime}$ S, 132$^{\circ}$ 42$^{\prime}$ 33$^{\prime \prime}$ E with a baseline established in a 230$^{\circ}$ SW/50$^{\circ}$ NE orientation from the origin point with a spacing of 20~m between points.  From each point in the baseline, five points in a 320$^{\circ}$ NW orientation and one in a 140$^{\circ}$ SE orientation were established with a 20~m line spacing to complete the total grid.  The final grid was 80~m by 120~m with a total of 30 grid points, which covered a majority of the depressionÕs floor (see Fig.~\ref{fig:2}) but did not encompass the entire feature or extend beyond the rim at any location, given our time constraints.

The gravity data were acquired using a Scintrex CG-3 gravitometer with base station readings made every hour to correct for instrument drift.  Drift, latitude, free air and Bouguer corrections were performed on the data using the dumpy-level topography for the free air and Bouguer corrections.  For the latter correction, we used a Bouguer density of 2400~kg~m$^{-3}$.  Modelling of the geophysical data was performed with the Encom software ModelVision 8.0, with the variable density structures discussed in Section~IV.

A geological survey, including fault mapping, was performed on the depression's walls.  15 strike and dip measurements (SD) were made along the depression's edge and rock samples (SV) were obtained at seven localities on the depression's outer edge, all of which are marked in Fig.~\ref{fig:2}.  Measurements taken were predominantly of the numerous near-vertical fractures found along the top of the depression edge, as well as the bedding planes of the sandstone unit hosting the depression.

The rock samples obtained were cut and polished into thin sections and examined for evidence of shocked quartz, as discussed in Section~V.  A cursory naked eye search of the area did not reveal any shatter cones near the wall's edge, although none were expected for a crater of this size.  No obvious meteorite fragments were visible, although we did not use a metal detector or other equipment to search the area.

\section{Morphology}

In terms of morphology, impact craters are generally divided into two categories: simple and complex, with simple craters (D~$<$~3.2~km) exhibiting a distinct bowl-shape with a high depth-to-diameter ratio ($\sim$1:5), while complex craters (D~$>$~3.2~km) exhibit a considerably flatter basin and feature a central uplift, with a much lower depth-to-diameter ratio ($\sim$10:1 or more) \cite{IEEEhowto:Melosh}\cite{IEEEhowto:Collins}.  With a diameter of $\sim$280~m, an impact origin would classify the Palm Valley depression as a simple crater.

\begin{figure}[h]
\centering
\includegraphics[width=6cm]{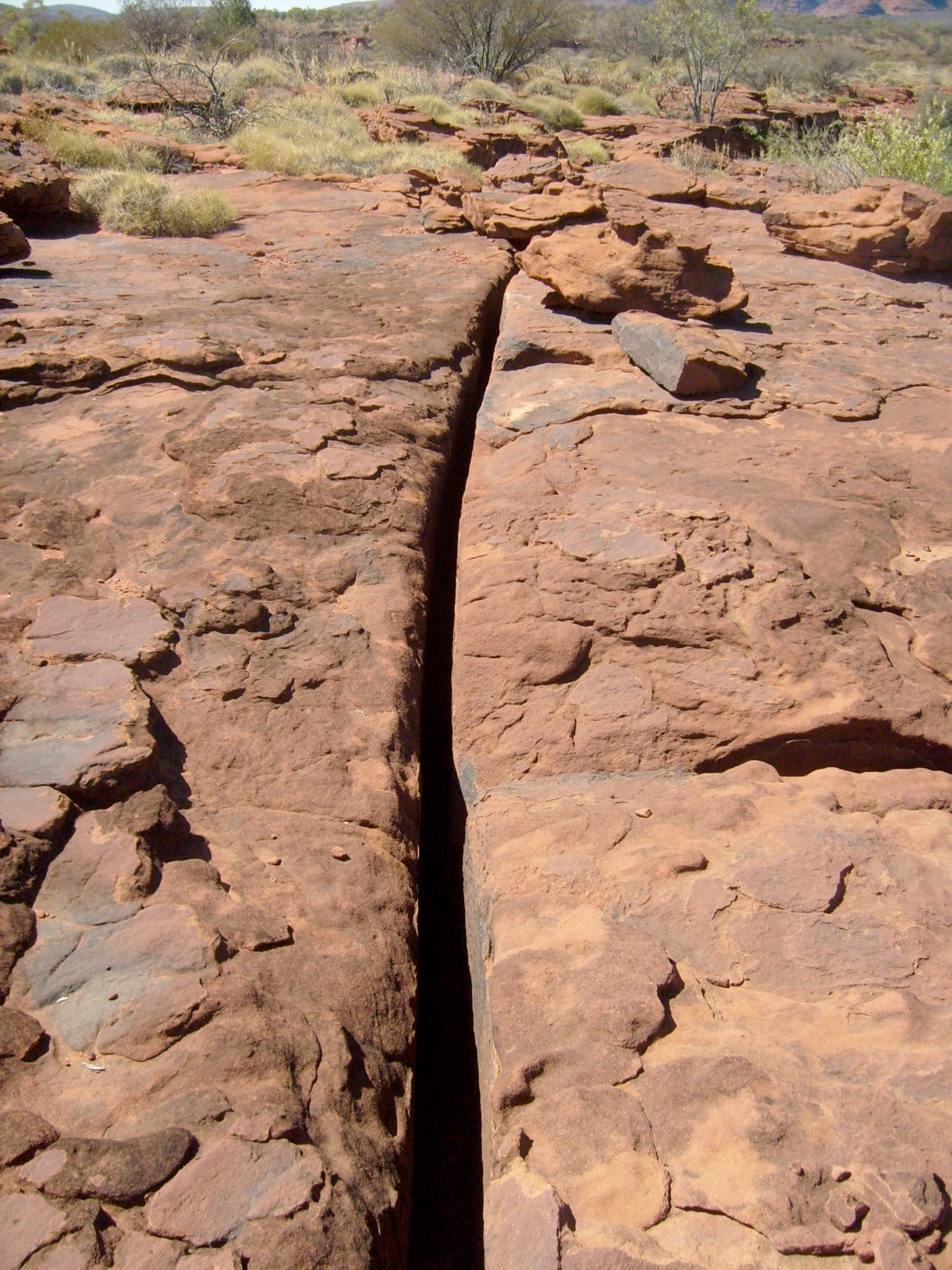}
\caption{A steeply dipping planar crack on the depression's edge, orientated tangential to trend of scarp.}
\label{fig:3}
\end{figure}

The Palm Valley depression exhibits a large, steep semicircular escarpment on the north side and a shallow wall on the south side with an outflow channel that has excised the depression's wall on the southeast corner.  No corresponding inflow channel exists, which indicates the outflow channel was formed from the drainage of water during rainfall events.  This is supported by the depression's topographical relief, which slopes towards this outflow channel (more recent Google Maps images after a wetter-than-normal season more clearly show the water outflow channel).  A large number of planar cracks and foliations are observed on the depression's wall, tangential to the strike of scarp, dipping near vertical (e.g. Fig.~\ref{fig:3}).  The depression appears to be heavily eroded, with the outer wall exhibiting large variations in relief ($>$~40~m) ascribed to differential erosion.

The height of the wall is lower on the southern side and the circularity of the depression is broken by the large slumping of rock to the east.  This has not been transported, suggesting the products of collapse are ineffectively removed.  Additionally, the outflow channel is small, resulting in difficulties in mass transport models of eroded material.  The existing Finke Gorge system demonstrates similar erosional features to those of the Palm Valley depression in terms of ultimate relief and rim or bank slope.  However, no inflow channel is observed and the northern side of the depression reveals an otherwise steeply sided wall.  The depression is also uniquely circular in symmetry, which is distinct from other erosional escarpments in the Finke Gorge system.

An erosional origin to this depression may be argued largely from its location in an existing gorge system and the similarities in slope and height in the steep-sided depression's walls with those of the adjoining gorge system.  On the other hand, the single outflow channel, sloping relief towards this channel, the apparent lack of buried fluvial channels, and the sub-parallel tangential fractures observed on the rim of the depression argue against an erosional origin.

In summation, the symmetrical circular, bowl-shaped morphology of the depression with a single outflow channel and no corresponding inflow channel is consistent with a meteorite impact.  However, Class~A diagnostic criteria are lacking.  No meteorite fragments were recovered and no shatter cones were observed near the depression's edge.  If the depression were the result of an impact, its morphology suggests the well-defined northern wall indicates the direction of the ejecta is to the north.  No ejecta blanket is observed, although a northerly blanket would be scoured by the Palm Valley gorge system.  Additionally, there is also no clearly defined uplift on the edge of the depression, indicative of a crater rim. 

\section{Gravity}

\begin{figure}[h]
\centering
\includegraphics[width=15cm]{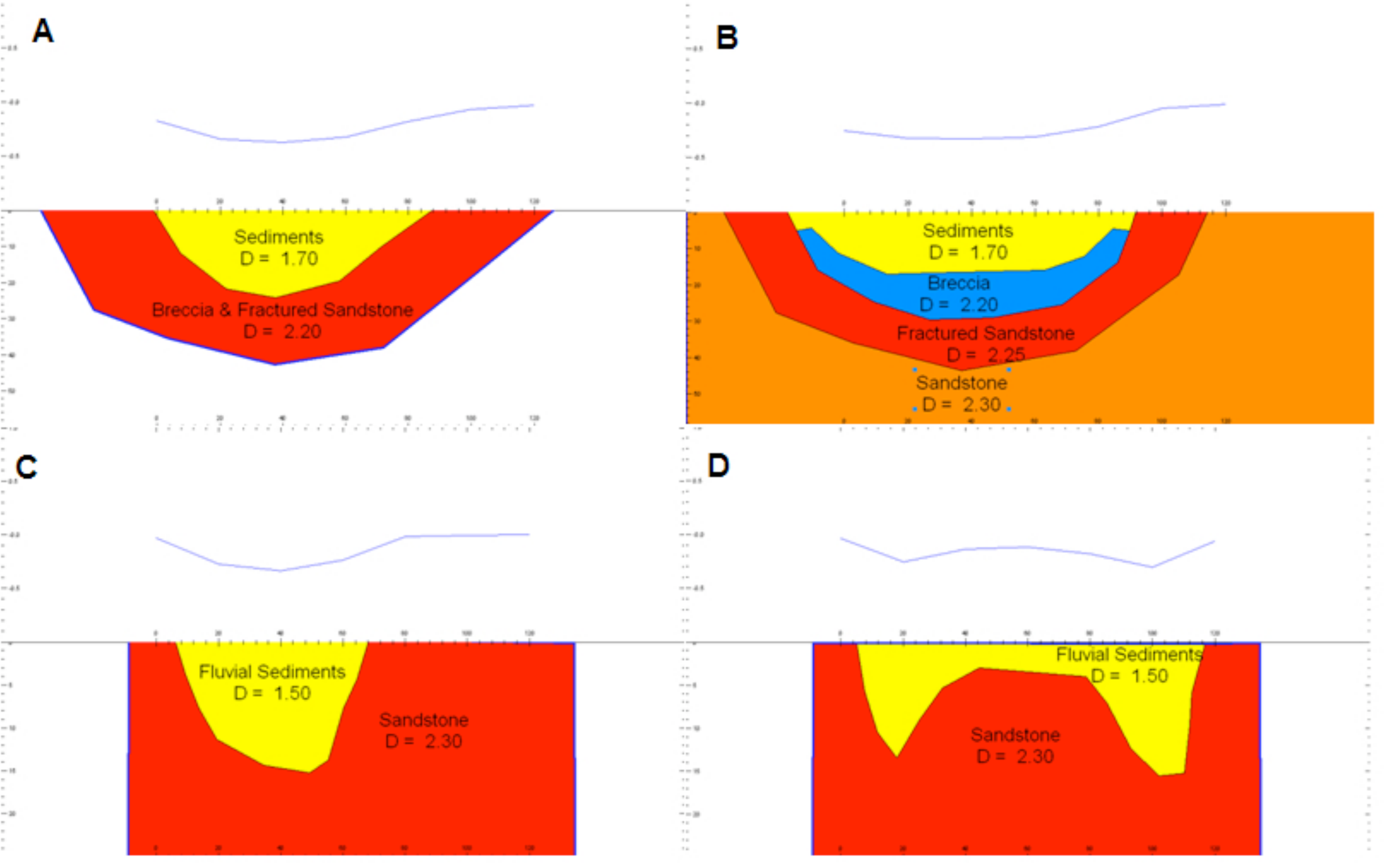}
\caption{Generic profiles and associated responses (blue line) of (A) 2-layer simple crater, (B) 4-layer simple crater, (C) fluvial channel, and (D) oxbow lake.}
\label{fig:4}
\end{figure}

We produce a number of data-free profiles of generic structures in order to compare to their modelled responses with those observed from the surveyed lines.  The two most probable formation mechanisms for the depression are either erosional or impact (no evidence of volcanism is found in the region and none is reported in the literature).  Both processes have fairly different implications for subsurface structure and gravity response of the depressionÕs floor.  To accomplish this, we model gravity responses of the subsurface structure for a generic buried fluvial channel, a section across a generic buried oxbow lake, and simulated simple impact craters, with both 2 and 4 layers (Fig.~\ref{fig:4}).  It should be noted that these profiles are not systematic, but rather indicative of the response of these subsurface structures for the given density contrasts.

\begin{figure}[h]
\centering
\includegraphics[width=15cm]{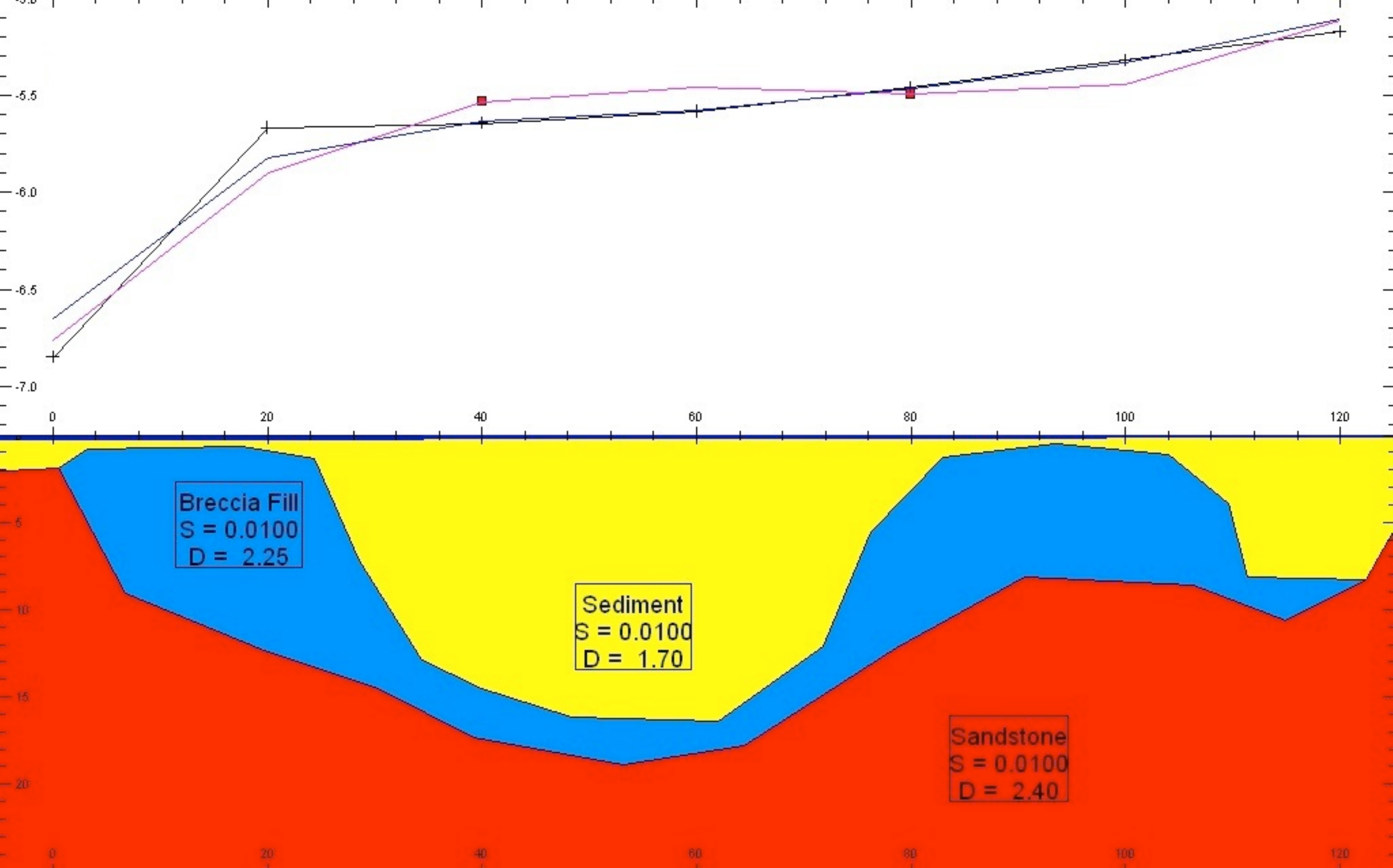}
\caption{Subsurface profile of Line 4 utilising Bouguer anomaly data, modelled with 3 layers (with D being the rock density). Shown are the reduced data (black lines), the modelled profile (blue), and the regional response (pink).  Profiles from other lines, in both 2 and 3 layers, show a similar trend.}
\label{fig:5}
\end{figure}

Subsurface modelling of Bouguer anomaly data was undertaken on Lines 2, 3 and 4, the latter of which can be seen in Fig.~\ref{fig:5}.  Initially, a two-layer model of the subsurface was attempted, comprising a basement layer of sandstone ($\rho$~=~2400~kg~m$^{-3}$) overlain by a sediment layer ($\rho$~=~1600-1700~kg~m$^{-3}$) with a thickness ranging from 1 to 12~m.  Although the modelled response fits the observed response reasonably well, the density of the sandstone and sediments is somewhat low since a fractured intermediate sandstone layer was not included.  As a result, a third layer was added to the model, primarily representing breccia infill (assuming an impact origin), with a density of 2250~kg~m$^{-3}$ and an average thickness of roughly 5$-$10~m, formed when the walls of the hypothetical transient crater collapsed and came to rest between the two modelled layers.  The newly modelled responses fit the observed responses with a high degree of accuracy and the structure conforms more closely to the expected structure of a simple impact crater than an oxbow lake or fluvial channel.  However, the layer of breccia fill added to the models could also be from processes not associated with the collapse of the transient crater wall, such as general fill from its surroundings over long periods of time.

One plausible explanation of the Palm Valley depression is erosion and/or collapse.  Such a process would be expected given the deeply dissected and eroded terrain of the region.  An erosional origin of the Palm Valley depression implies a deep, buried fluvial channel proximal to the depression's wall.  Given its circular shape, this channel should have been apparent twice on a transecting gravity profile, representing both an inflow and outflow channel.  However, the response expected from a buried channel is not seen (Fig.~\ref{fig:5}).

No volcanics were observed near the depression and none are reported from the region in the literature.  Additionally, the gravity response in Fig.~\ref{fig:5} precludes the presence of volcanics at depth.

\section{Mineralogy}

The presence of planar fractures (PFs) or planar deformation features (PDFs) in quartz grains could prove an impact origin.  However, at $\sim$280~m in diameter, the average peak pressure at the rim resulting from an impact should not have exceeded 7.5~GPa, the limit required for the formation of PDFs \cite{IEEEhowto:Melosh}.  Non-linearities in the propagation of the shock front could have lead to focussed stress at the grain scale, and thus shock deformation features could be found in a very small number of grains where, locally, the grain-scale pressure exceeded 7.5~GPa.  Low-level shock waves (possibly $<$~5~GPa, but definitely $<$~10~GPa) can form PFs in a small percentage of grains, but these features are not ubiquitous.

\begin{figure}[h]
\centering
\includegraphics[width=16cm]{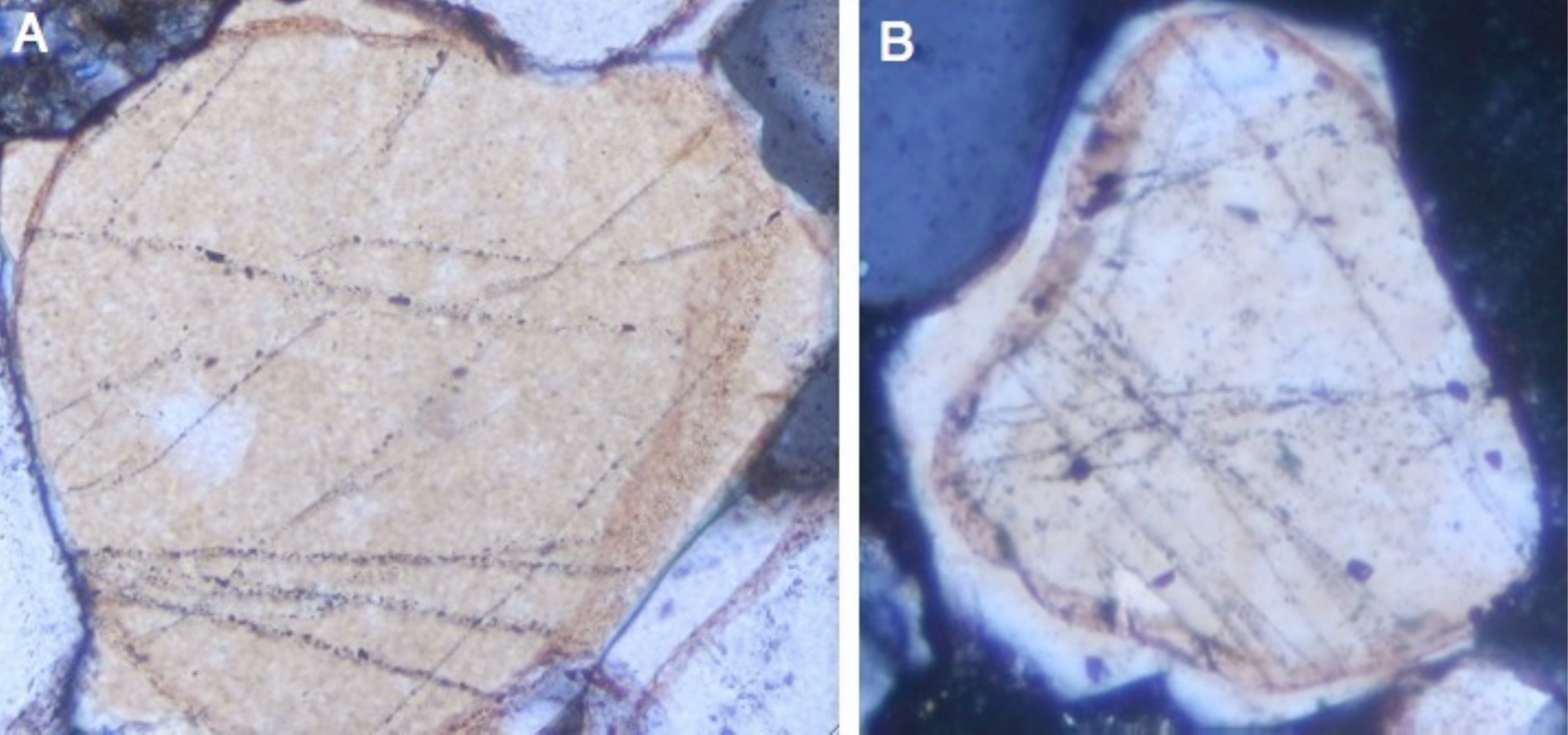}
\caption{Fracture features found in quartz grains, shown in cross-polarised light.  These images show truncation of the fractures by the cement, suggesting they may be older than the cement and not related to this impact.  The grains ($\sim$0.5~mm in width) were taken from PV1 (A) and PV2 (B).  See Fig.~\ref{fig:2} for PV locations.}
\label{fig:6}
\end{figure}

Microscopic analysis of the thin sections shows the rock consists predominantly of quartz ($\sim$45\%) and plagioclase feldspar ($\sim$38\%) mineral grains, $\sim$2-3\% microcline, and the remainder consisting of opaque minerals, in particular hematite.  Although individual grains are well rounded, the rock is poorly sorted, with sizes ranging from 10-200~$\mu$m.  Iron staining of mineral grains is readily observable under plain polarised light as an orange to red halo around individual grains, giving the rock a red to red-brown hue when viewed macroscopically.   Out of seven thin sections, four quartz grains revealed fracture features (Fig.~\ref{fig:6}) that appear sub-parallel, regular, and planar.  These are unlikely to be Boehm lamellae as the porosity of the rock is quite high, precluded the high-strain tectonism generally associated with such features, and the unit is tectonically undeformed, situated in a stable shield.

The thin sections do not show PDFs and the fracture features could have been formed by terrestrial processes, such as cracking during diagenesis.  Radiating fractures from a local focus of pressure (e.g. adjacent grain boundary) are observed, which are reminiscent of concussion fractures produced by low-pressure shock waves \cite{IEEEhowto:French}.  Concussion fractures have been used to support an impact origin in other structures \cite{IEEEhowto:Milstein88}\cite{IEEEhowto:Stone}\cite{IEEEhowto:Milstein01} but are not in themselves a diagnostic.  However, the fracture features terminate at the grain boundaries and do not cross into the cement overgrowths, indicating the cement post-dates the fractures.  This suggests the fractures are a feature of the original rock that contained the quartz.  Inclusions are also abundant in the rounded interiors of the grains and are absent in the clear overgrowths.  Therefore, it is unlikely that these fractures are related to this impact.

No evidence of melt is found in the thin sections and none is expected given the small size of the depression.

\section{Age Estimate}

Given the low erosion rate of the Finke River system, the heavily eroded nature of the depression suggests that if it is of impact origin, it is quite old.  It is difficult to constrain the depressionÕs age with certainty, as the average erosion rate over the lifetime of the river system has probably varied significantly, and is currently very low.  The deeply incised meanders of the Finke River system would have been created at a time of significant regional uplift.  Fission track data \cite{IEEEhowto:Kohn} date the last significant uplift in the Amadeus Basin region to between 150-450~Ma.  Assuming the Palm Valley Gorge system, with typical relief of $\sim$60~m, has conservatively eroded over a 150-450~Myr time period, the erosion rate would be between 0.13-0.40~m~Myr$^{-1}$.  Assuming the depression is a crater, a scaling relationship between the rim height relative to the level of the surrounding plains (h$_{\rm rim}$) and the crater diameter (D), in meters \cite{IEEEhowto:Melosh}, is given as

\begin{equation}
{\rm h_{rim} = 0.036~D^{1.014}}
\end{equation}

With an approximate diameter of 280~m, h$_{\rm rim}$~is approximately 11~m.  Given the differential erosion of the rim (non-existent at the southern outflow, to almost uneroded to the north, which drops sharply to a gorge), calculating the average amount of erosion on the rim is again difficult.  If we assume the southern (completely eroded) rim has eroded down from 11~m elevation, the estimated erosion rates provide an age of 27-85~Ma.  This sort of estimate is obviously fraught with enormous uncertainties, and until exposure ages are calculated, the age may date from 10 to 100~Ma or more. 

\section{Discussion}

An erosional origin of the depression is argued against based on its morphology and the gravity data.  However, we recognise that no Class~A diagnostic indicators are found.  Shatter cones, PDFs, and high pressure glass or melt are not expected given the depression's small size.  Although radiating fractures from local foci of pressure are observed in the quartz grains (similar to concussion fractures produced by low-pressure shock waves), they are not considered shocked quartz and are insufficient as a diagnostic.

With the exception of meteorite fragments, most of the diagnostics required to identify impact structures \cite{IEEEhowto:French} are biased toward larger impacts (D~$>$~500~m).  Applying these same diagnostic criteria to smaller impacts, especially those that are quite old, is inadequate.  Very small craters (D~$<$~300~m), which should outnumber larger impacts, are generally unable to produce most diagnostic indicators (e.g. shatter cones, shocked quartz, high pressure glass or melt, etc), relying almost solely on the presence of meteorite fragments to confirm an impact origin.  For example, shatter cones at Wolfe Creek crater (D~$\approx$~880~m) are rare \cite{IEEEhowto:ONeill} and none were found at Henbury (150~m), Dalgaranga (20~m), Boxhole (170~m), Veevers (80~m), or the probable Hickman crater (270~m).  Additionally, the energy from smaller impacts is generally insufficient to produce PDFs in shocked grains.  None were found at Henbury, Boxhole, Veevers, Dalgaranga, or Hickman and were rare ($\sim$2-3\%) in samples from Wolfe Creek \cite{IEEEhowto:ONeill}.

The single diagnostic concluding an impact origin for Boxhole, Dalgaranga, Henbury, and Veevers is the presence of meteorite fragments, which exist primarily because of their geologically young ages ($<$~0.5~Ma) \cite{IEEEhowto:Shoemaker90}\cite{IEEEhowto:Shoemaker05}.  For older craters ($>$~0.5~Ma), meteorite fragments would have mostly or completely eroded away leaving no other diagnostics.  

These limited diagnostics highlight a need for developing recognition criteria to help identify small, ancient impact craters where Class~A indicators would either not have been present due to their small size or would have been erased by the passage of time.  We also encourage systematic searches for impact structures in Central Australia, as our current knowledge of the meteoroid influx rate combined with the relatively low number of confirmed impact structures and the antiquity of the Central Australian landscape suggests many more craters remain to be discovered.

\section{Conclusion}
Morphological, geophysical, and mineralogical data from a survey suggests the Palm Valley depression was not formed by erosional processes.  The authors consider an impact origin as a preferred explanation.  The lack of diagnostics preclude a positive identification as a crater, although its small size and antiquity may be the reason behind this.  We encourage further discussion about the depression's origins and the development of diagnostic indicators for small, eroded craters in general.




\section*{Acknowledgments}

We wish to thank the Departments of Earth \& Planetary Science and Indigenous Studies at Macquarie University and Ray Norris.

Palm Valley is a jointly managed park, established under the {\it Territory Parks \& Wildlife Conservation Act}.  The management partners (Northern Territory Parks \& Wildlife Service and the Traditional Owners) jointly develop guidelines on the environmental and cultural management of the park.  Any access or activity to be undertaken outside of the normal tourist areas require a permit from the Parks \& Wildlife Service who will consult with the joint management committee prior to issuing or refusing a permit.



%

\end{document}